\input{aipcheck}

\documentclass[
    ,final            
  ]
  {aipproc}

\layoutstyle{8x11single}

\begin{document}

\title{MicroBooNE:\newline
A New Liquid Argon Time Projection Chamber Experiment}


\author{Mitchell Soderberg, for the MicroBooNE Collaboration}{
  address={Yale University, Dept. of Physics, P.O. Box 208120, New Haven, CT  06520, USA}
}

\begin{abstract}
Liquid Argon Time Projection Chamber detectors are well suited to study neutrino interactions, and are an intriguing option for future massive detectors capable of measuring the parameters that characterize neutrino oscillations.  These detectors combine fine-grained tracking with calorimetry, allowing for excellent imaging and particle identification ability.  In this talk the details of the MicroBooNE experiment, a 175 ton LArTPC which will be exposed to Fermilab's Booster Neutrino Beamline starting in 2011, will be presented.  The ability of MicroBooNE to differentiate electrons from photons gives the experiment unique capabilities in low energy neutrino interaction measurements. 
\end{abstract}

\maketitle


\section{Introduction}
Liquid Argon Time Projection Chambers (LArTPCs) are an appealing class of detectors that offers exceptional opportunities for studying neutrino interactions thanks to the bubble-chamber quality images they can provide.  The unique combination of position resolution, calorimetry, and scalability provided by LArTPCs make them a desirable technology choice for future massive detectors.  There is an active program in the U.S. to develop LArTPCs, with the final goal of constructing a massive detector that can be used as a far detector in a long-baseline neutrino oscillation experiment.  The MicroBooNE experiment is the next step in this program, and it will be the focus of this document.

\section{LArTPC Technique}
The LArTPC technique has been around for several decades, with pioneering work done as part of the ICARUS experiment \cite{Rubbia, ICARUS}.  A wire chamber is placed in highly-purified liquid Argon, and an electric field is created within this detector.  Neutrino interactions with the Argon inside the detector volume produce ionization electrons that drift along the electric field until they reach finely segmented and instrumented anodes ($\textit{i.e.}$ - wireplanes) upon which they produce signals that are utilized for imaging and analyzing the event that occurred.  Applying proper bias voltages to the wireplanes, such that electrons drift undisturbed through the initial planes, allows several complementary views of the same interaction which provide the information necessary for three-dimensional event imaging\cite{Grids}.  Calorimetric measurements can be extracted from the pulses that are observed on the wireplanes.

This technique allows for very precise imaging, the resolution being dependent on several factors:  wire pitch, plane spacing, sampling rate, and electronics S/N levels.  The wire pitch is typically on the order of several millimeters, and the rapid sampling rate ($\approx$5MHz) typical of these detectors equates to an image resolution of fractions of a millimeter along the drift direction (which is the coordinate common to all the wireplanes of the TPC).  The technology is further made attractive in that the number of electronics channels required for the detector does not scale directly with the volume of the detector.   This scaling feature, along with the relatively low cost of Argon, makes LArTPCs an intriguing option for future massive neutrino detectors.

  \begin{figure}[!h] 
   \centering
   \includegraphics[height=2.75in,width=3in]{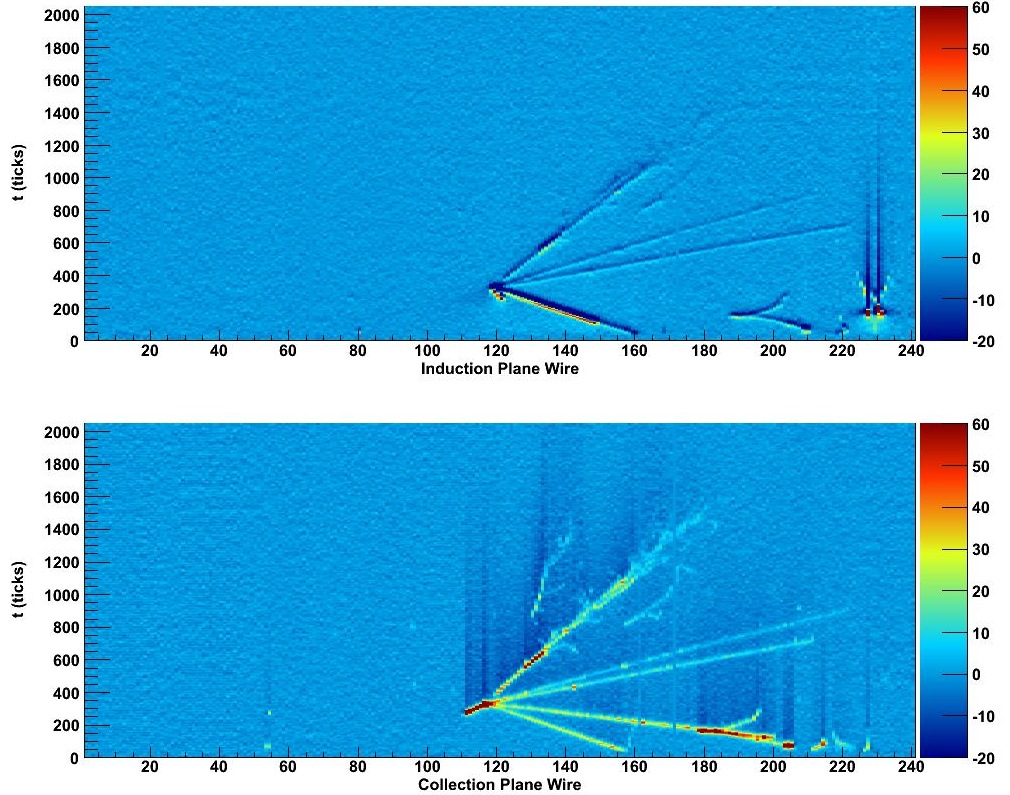}  
   \caption{Neutrino event from ArgoNeuT, a LArTPC that operates in the NuMI beamline at Fermilab.}
   \label{fig:argoneut}
\end{figure}

While LArTPCs are an intriguing detector technology, they are not without their challenges.  One of the biggest challenges is producing and maintaining Argon that is pure enough to allow the ionization to drift for the required distances.  To address this issue, several U.S. institutions have worked on developing and testing new filters that can cleanse the Argon to the required levels necessary for a LArTPC experiment, and can also be regenerated when they have become saturated \cite{Filter}.  These new filters are a necessary step along the path to massive detectors, and they have already been used by several test stands built in the U.S. with the goals of studying detector material effects on Argon purity, and looking for cosmic-ray events in an LArTPC \cite{Yale}.  These filters are also used by the ArgoNeuT detector, which is a small ($\approx$175 liter) LArTPC currently taking data in the NuMI neutrino beam at Fermilab, an event from which is shown in Figure \ref{fig:argoneut} to illustrate the exceptional imaging possible with these detectors\cite{ArgoNeuT}.

\section{MicroBooNE}
\begin{figure}[htbp] 
   \centering
   \includegraphics[height=2.75in,width=3in]{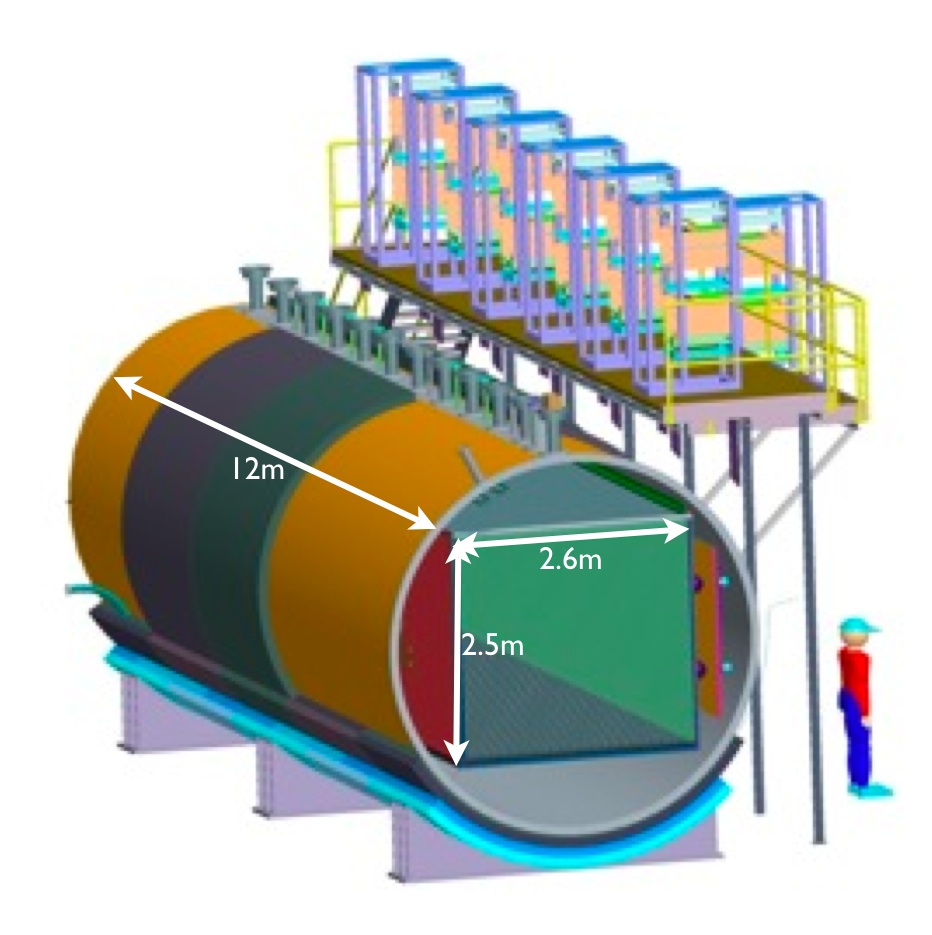}  
   \caption{Diagram of MicroBooNE experiment.}
   \label{fig:microboone}
\end{figure}

The next crucial step in the U.S. LArTPC program is MicroBooNE, an experiment with Stage 1 approval at Fermilab that is partially funded by the NSF.  MicroBooNE will answer important physics questions and will also provide useful hardware and software development for future massive ($\geq$5 ktons) detectors.  The experiment consists of a 2.5mx2.5mx12.0m TPC containing $\approx$90 tons of liquid Argon, that will be placed in a cryostat as depicted in Figure \ref{fig:microboone}.  The TPC  will be instrumented with three wireplanes, each with 3mm wire pitch, totaling approximately 10000 channels of electronic readout.  This experiment will be positioned on the surface at Fermilab, and will be simultaneously exposed to the on-axis Booster Neutrino Beamline (BNB), and the off-axis NuMI beamline.  MicroBooNE will utilize 30 cryogenic photomultiplier tubes (PMTs) that will capture the abundant scintillation light present in neutrino interactions with Argon.  The PMTs will provide trigger signals that can be required in coincidence with the beam arrival in order to eliminate empty spills and reduce the data output of the experiment.

MicroBooNE will have several features that provide valuable input for future massive LArTPC detectors.  For example, there is a two phase electronics program planned for MicroBooNE that will evolve the preamplifiers from JFET to CMOS technology.  In the initial phase MicroBooNE will utilize JFET preamplifiers and place them inside of the cryostat, in an 8$\%$ gas ullage at the top of the cryostat.  The cold temperature ($\approx$120K) of the ullage gas will allow the preamplifiers to operate with a S/N that is a factor of 3 better than that at room temperature.  The second phase refers to the development of a custom CMOS preamplifier that can be placed in the liquid, directly on the TPC.  Placing the electronics inside the cryostat is likely to be a requirement for future massive detectors that will have to combat noise incurred during the long transit a signal, that may already be diminished after drifting many meters, must make up and out of the very large detector.  By amplifying the signal close to the source, noise levels in the data can be kept to a minimum.  Placing the preamplifiers inside of the detector introduces several challenges that must be considered in the design of the detector.  Understanding the heat load the preamplifiers will place on the cryogenic system and also the impact they will have on the Argon purity are among the challenges that must be addressed.

MicroBooNE will have a maximum drift distance of $\approx$2.5m, which is significantly longer than any existing LArTPC.  Operating with this long drift will provide useful input for future massive detectors where the drift distance might be equivalent or even longer ($\textit{e.g.}$ - $\geq$5m).  At an operating electric field of 500V/cm (corresponding to a drift velocity of 1.6mm/$\mu$s), MicroBooNE will have to sample for 1.5ms in order to capture the entirety of events occurring throughout the full drift distance of the TPC.  In fact, MicroBooNE will most likely sample a full drift-time before and after the window associated with the beam, to aid in eliminating tracks not ssociated with the beam (\textit{i.e.} - from copious cosmic ray events that penetrate into the detector) that can overlap with the neutrino event.  The long drift distance designed into the TPC imposes demanding requirements on Argon purity, and also on S/N levels in the electronics.  Efforts to reconstruct events in the TPC will have to address the effects of electron recombination and diffusion that can occur over such long drift distances.

Another interesting feature of the MicroBooNE experiment that will be important for future detectors is a purity demonstration it will carry out, filling the cryostat with liquid after an initial gaseous purge, as opposed to filling after an initial vacuum pumpout.  The goal of this demonstration is to determine whether or not the required argon purity level can be achieved in a fully instrumented detector starting from a non-evacuated environment.  This is crucial information needed in the design of massive detectors, which may be so large that they cannot be safely evacuated due to structural risks.

\section{MicroBooNE Physics}

\begin{figure}[htbp] 
   \centering
   \includegraphics[height=2.75in,width=3in]{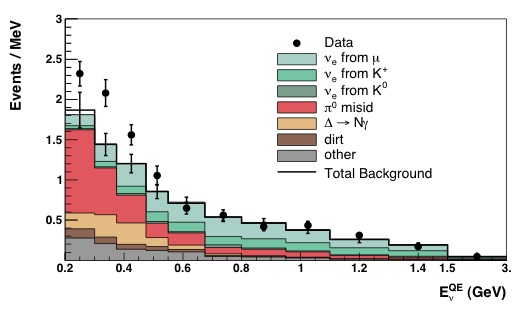}  
   \caption{Excess observed in low-energy region of MiniBooNE $\nu_{e}$ appearance analysis\cite{Miniboone}.} 
   \label{fig:miniboone}
\end{figure}

MicroBooNE will perform a wide range of physics analyses during its 2-3 year run.  A primary goal is to examine the excess of low-energy $\nu_{e}$ appearance events observed by the MiniBooNE experiment, as shown in Figure \ref{fig:miniboone}\cite{Miniboone}.  MiniBooNE can not easily differentiate electrons from photons so it is unable to determine the nature of the events responsible for the excess.  LArTPCs are very good at differentiating electrons from photons, so MicroBooNE plans to conclusively identify the nature of the low-energy excess.  The main method of separating electrons and gammas is to calculate the average energy deposition, or $dE/dx$, along the first few centimeters of a candidate track.  Tracks originating from electrons will tend to cluster around 1 MIP deposition, while tracks originating from photons will cluster around 2 MIP deposition after the photon pair converts.  It is thought that $dE/dx$ alone will allow better than 90$\%$ efficiency at separating out electrons and gammas, and this efficiency may be improved further if other topological cuts are folded into the analysis.  The excellent particle identification capability afforded by this $dE/dx$ technique means that certain backgrounds in the MiniBooNE $\nu_e$ analysis, such as neutral-current pion (NC $\pi^{\circ}$) events where the gammas from the pion decay fake an electron, can be almost completely eliminated.

MicroBooNE will make important cross-section measurements that are useful for future massive LArTPCs and are also particularly relevant to the neutrino oscillation community because of the low-energy ($\leq$1GeV) of the neutrinos in the BNB.  The most important events for oscillation physics are charged-current quasi-elastic (CCQE) scattering events, which form the signal sample in oscillation analyses.  The CCQE sample is also interesting in that it can be used to measure the axial-vector form factor, possibly for the first time on Argon, and gives some insight into nuclear effects.  MicroBooNE should obtain 68k $\nu_{\mu}$ CCQE and 500 $\nu_{e}$ CCQE events in the requested running time in the BNB.  MicroBooNE will also collect around 8k NC$\pi^{\circ}$ events from the BNB which will be very useful in developing the $dE/dx$ technique and reconstructing pions.  MicroBooNE should also collect a small ($\approx$500) sample of charged kaons that will also be useful in developing the $dE/dx$ technique and understanding the potential of a future massive LArTPC to search for proton decay via the $p\rightarrow \overline{\nu}K^{+}$ channel.

MicroBooNE will have the capability to record supernova neutrinos, but not without some help.  MicroBooNE will not have a supernova trigger of its own, rather it will buffer one hour of data on removable storage devices and wait for communication from supernova watch networks.  If a supernova is detected and the alert relayed within this hour buffer time, then MicroBooNE can use the information recorded on the storage device to look for supernova neutrinos.  Due to the small size of MicroBooNE only a few dozen events would be expected from a galactic supernova.  Several of the interaction channels for supernova neutrinos, such as CC absorption or NC events, will require sensitivity to very low-energy photons (<50MeV), so studies of the detection thresholds of the PMT system will be necessary to understand the feasibility of this analysis.

MicroBooNE may attempt to measure the strange quark content of the proton, $\Delta_{s}$, through ratio measurements of NC ($\nu_{\mu}p\rightarrow\nu_{\mu}p$)and CC ($\nu_{\mu}n\rightarrow\mu^{-}p$) scattering.  Measuring this ratio as a function of the momentum transfer ($Q^2$) of the interaction provides a handle on $\Delta_{s}$.  The combination of low-energy neutrinos from the BNB, and the low-energy thresholds MicroBooNE is capable of, indicate that this ratio could be measured down to very low values of $Q^{2}$ where the sensitivity to $\Delta_{s}$ is strongest.  The main challenge of this measurement will be estimating the backgrounds from neutral-current neutron events ($\nu_{\mu}n\rightarrow\nu_{\mu}n$), which can fake the signature of neutral current proton events if the final-state neutron liberates a proton from an Argon nucleus.

To aid in studies of MicroBooNE physics potential, and as a necessary step for the evolution of LAr TPC technology, the collaboration is developing a complete software analysis environment.  This is a joint effort between members of MicroBooNE, ArgoNeuT, and any other future LAr TPC experiment (such as massive detectors in a long-baseline scenario) to develop a general purpose set of tools that can be used once the geometrical specifics of the detector are provided.  The collaboration is using the FMWK code environment to develop tools for simulating and reconstructing neutrino interactions in liquid Argon \cite{FMWK}.  A new GEANT4 based simulation is being developed that can take neutrino interaction information from an outside source (such as GENIE, NUANCE, etc...) and simulate the evolution of the event in liquid Argon as ionization is created and drifted to wireplanes, and as pulses develop and are processed by the electronics.  Reconstruction tools are being written that will identify the interesting portions of an electronics channel readout, and then use these portions to create a three-dimensional view of the interaction complete with calorimetric information.  Automated reconstruction of events is the final goal of this effort, since visually scanning the hundreds of thousands of events produced by MicroBooNE (and the much greater statistics that a $\geq$5kton detector would produce) is not a viable strategy.  The possibility of testing software tools on neutrino data from the ArgoNeuT project will help tremendously in preparing for MicroBooNE.

\section{Conclusion}

There is a very active program of LArTPC development in the U.S., geared towards the goal of building a massive detector to study neutrino oscillation physics.  Several test projects already exist in the U.S. to study Argon purity, and to collect a sizable sample of neutrino interactions that can be utilized to develop software tools.  MicroBooNE is the next step in the U.S. for LArTPC development, and aside from performing many interesting physics analyses, it will help answer several hardware questions that are important for future massive detectors.

\bibliographystyle{aipproc}   



\begin{thebibliography}{9}

\bibitem{Rubbia}The Liquid-argon time projection chamber: a new concept for Neutrino Detector, C. Rubbia, CERN-EP/77-08 (1977);
\bibitem{ICARUS}Design, construction and tests of the ICARUS T600 detector, Nucl. Inst. Meth., A527 329-410 (2004);
\bibitem{Grids}Design of Grid Ionization Chambers, O. Bunemann, T.E. Cranshaw, and J.A. Harvey; Canadian Journal of Research, 27, 191-206, (1949);
\bibitem{Filter}A Regenerable Filter for Liquid Argon Purification, A. Curioni \textit{et. al}, Nucl. Inst. Meth., A605 306-311 (2009);
\bibitem{Yale}The Yale Liquid Argon Time Projection Chamber, A. Curioni, B. Fleming, M. Soderberg, arXiv:0804.0415 (2008);
\bibitem{ArgoNeuT}t962.fnal.gov
\bibitem{Miniboone}Unexplained Excess of Electron-Like Events From a 1-GeV Neutrino Beam, A.A. Aguilar-Arevalo \textit{et. al.}, Phys. Rev. Lett. 102,101802 (2009);
\bibitem{FMWK}http://enrico1.physics.indiana.edu/fmwk/wiki

\end{thebibliography}



\end{document}